\documentclass[twoside,12pt]{article}
\usepackage{epsfig}
\usepackage{graphicx}
\usepackage{amsmath}

\newcommand{\be}{\begin{equation}}
\newcommand{\ee}{\end{equation}}
\newcommand{\bea}{\begin{eqnarray}}
\newcommand{\eea}{\end{eqnarray}}

\newcommand{\bold}[1]{\mbox{\boldmath ${#1}$}}
\newcommand{\fslash}[1]{\mbox{$\!\not\!#1$}}
\begin{document}

\title{ \vspace{1cm} Polarized structure functions of nucleons and nuclei}
\author{W.\ Bentz,$^{1}$ I. C.\ Clo\"et,$^{2}$ T.\ Ito,$^1$ 
A. W.\ Thomas,$^3$ K.\ Yazaki$^4$\\
\\
$^1$Department of Physics, School of Science, Tokai University, \\ 
Hiratsuka-shi, Kanagawa 259-1292, Japan\\
$^2$Physics Division, Argonne National Laboratory, \\
Argonne, IL 60439, USA\\
$^3$Jefferson Lab, 12000 Jefferson Avenue, Newport News, \\
VA 23606, USA\\
$^4$Department of Physics, Tokyo Woman's Christian University, \\
Suginami-ku, Tokyo 167-8585, Japan}
\maketitle
\begin{abstract} 
We determine the quark distributions and structure functions for both
unpolarized and polarized DIS of leptons on nucleons and nuclei. The
scalar and vector mean fields in the nucleus modify the motion of
the quarks inside the nucleons. By taking into account this
medium modification, we are able to reproduce the experimental data on the
unpolarized EMC effect, and to make predictions for the polarized
EMC effect. We discuss examples of nuclei where the polarized EMC
effect could be measured. We finally present an extension of our
model to describe fragmentation functions.
\end{abstract}
\section{Introduction}

In order to describe nonperturbative effects like spontaneous
chiral symmetry breaking and nuclear binding on the level of quarks, 
effective chiral quark theories are powerful tools.
A prominent example is the EMC effect, which has clearly shown that the
quark distributions of bound nucleons differ
from those of free nucleons \cite{Arneodo:1992wf}. It has been shown recently \cite{Cloet:2006bq} 
that this
effect can be explained if one takes into account the response of
the quark wave function to the nuclear environment, that is, to the nuclear
mean fields, and that the same mechanism gives rise also to medium 
modifications of the polarized quark distributions and structure 
functions. In this work, we will focus on the predictions for the
polarized EMC effect, and briefly discuss extensions to describe
transversity distributions \cite{Cloet:2007em} and fragmentation 
functions \cite{progress}.  

\section{Quark distributions and structure functions}

In this work we will mainly be concerned with the following 
EMC ratios:
\begin{eqnarray}
R(x) &=& \frac{F_{2A}(x_A)}{Z F_{2p}(x) + N F_{2n}(x)} \,,  \nonumber \\
R_s^H(x) &=& \frac{g^H_{1A}(x_A)}{P_p^H g_{1p}(x) + P_n^H g_{1n}(x)} \,.  
\label{ratios} 
\end{eqnarray}
Here $x$ is the Bjorken variable for the nucleon, and $x_A$ is $A$ times
the Bjorken variable for the nucleus of mass number $A$, so that
$0<x_A<A$. The unpolarized and polarized structure functions of the
nucleon are denoted as $F_{2 \alpha}$ and $g_{1 \alpha}$ respectively
($\alpha=p,n$), while $F_{2 A}$ and $g^H_{1 A}$ are the corresponding
structure functions of the nucleus $A$ with spin projection 
$H=-J, \dots +J$ along 
the direction of the incoming electron. The polarization factors of
protons and neutrons, which appear in the denominator of the spin
dependent EMC ratio, are defined as twice the expectation values of 
the proton and neutron spin operators between the polarized nuclear
states. Both ratios in Eq.~\eqref{ratios} are defined so that they
become unity in a naive single particle model based on nonrelativistic
nucleons.

The parton model expressions for the nuclear structure functions
are very similar to those of the nucleon \cite{Jaffe:1988up}, for example    
\begin{eqnarray}
g_{1A}^H(x_A) = \frac{1}{2} \sum_q e_q^2 \, \Delta q_A^H(x_A)
= \frac{1}{2} \sum_q e_q^2 \left( q_{A \uparrow}^H(x_A) -
q_{A \downarrow}^H(x_A) \right) \, . \label{part}
\end{eqnarray}  
Here $q_{A \uparrow}^H(x_A)$ is the probability to find a quark with
momentum fraction $x_A/A$ and $s_z=1/2$ in the nucleus $A$ with
$J_z=H$, and similar for $q_{A \downarrow}^H(x_A)$.

It is important to keep in mind the following two points:
First, usually only a few valence nucleons (or holes) contribute
to the nuclear polarization, and therefore $g_{1A}^H$ is of order
$1/A$ relative to $F_{2A}$, where all nucleons contribute. Second,
the structure function of a free proton is larger and much better
known than the neutron structure function. Therefore, possible
candidates for the observation of the polarized EMC effect are
stable nuclei which are not too heavy, and where the polarization 
is dominated mainly by the protons.

Nuclear spin sums are interesting quantities, which have not yet 
been explored in detail. The isoscalar and isovector combinations are
\begin{eqnarray}
\int {\rm d}x_A \, \left( \Delta u_A^J(x_A) + \Delta d_A^J(x_A) \right)
&=& \Sigma \, (P_p^J + P_n^J) \,, \nonumber \\ 
\int {\rm d}x_A \, \left( \Delta u_A^J(x_A) - \Delta d_A^J(x_A) \right)
&=& g_A \, (P_p^J - P_n^J)\,,  \label{sum}
\end{eqnarray}
where we assumed for simplicity that only
a single nucleon state contributes to the nuclear polarization. 
The first relation contains information on the quark spin sum in a 
bound nucleon ($\Sigma$), and the second one on the axial coupling constant
of a bound nucleon ($g_A$). The latter is related
to nuclear Gamow-Teller matrix elements \cite{Arima:1988xa}, and establishes an 
important link between quark physics and nuclear structure physics.

\section{Model calculations}

We describe the nucleon as a bound state of a quark and a diquark
by using the Faddeev framework in the Nambu--Jona-Lasinio (NJL) model.
Although the relativistic Faddeev equation, which is represented
graphically in Fig.~1, can be solved exactly 
in the NJL model \cite{Ishii:1995bu}, for our applications to nuclei we 
limit ourselves
to the static approximation, where the momentum dependence of the 
quark exchange kernel is neglected \cite{Buck:1992wz}. We take into
account the scalar and axial-vector diquark channels, and avoid
unphysical quark decay thresholds by introducing an infrared cut-off,
in addition to the ultraviolet one, in the proper-time regularization
scheme \cite{Bentz:2001vc}. 

\begin{figure}[tb]
\centering\includegraphics[scale=1.6]{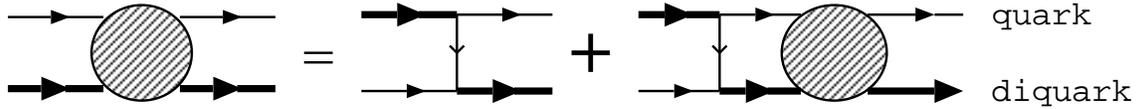}
\begin{minipage}[t]{16.5 cm}
\caption{Graphical representation of the Faddeev equation. The single
line denotes a quark, and the bold line a diquark.\label{fig1}}
\end{minipage}
\end{figure}

By using the quark-diquark vertex functions, we 
evaluate the Feynman diagrams shown in Fig.~2 with the operator
insertions $(\gamma_-, \gamma_- \gamma_5,
\gamma_- \gamma^1 \gamma_5) \times \delta(x - k_-/p_-)$,
to get the unpolarized, polarized, and transversity
quark distributions, respectively. Here $k$ and $p$ are the momenta of the quark and the nucleon, 
and $a_-$ denotes the light-cone minus-component of a 4-vector $a$. 
It is straight forward to extend this calculation to the case of a 
bound nucleon by including the mean nuclear scalar and vector fields 
into the quark propagators of Fig.~2.

\begin{figure}[tb]
\centering\includegraphics[scale=0.8]{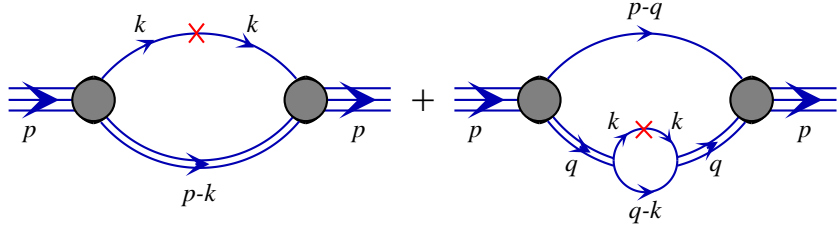}
\begin{minipage}[t]{16.5 cm}
\caption{Feynman diagrams for the quark distribution functions in the
nucleon. The operators inserted into the quark lines are explained
in the text. \label{fig2}}
\end{minipage}
\end{figure}

Finite nuclei in our present approach are described in a
simple independent particle picture: 
We assume Woods-Saxon scalar and vector potentials for nucleons,
with depth parameters determined from our earlier self-consistent
nuclear matter calculations \cite{Bentz:2001vc}, and standard values for the range
and diffuseness parameters \cite{bm}. After solving the Dirac equation with
these potentials, we calculate the expectation values of the
scalar and vector potentials for each nucleon orbit, and use the
quark-diquark (Faddeev) equation to translate them into the average 
scalar and vector fields for quarks. These
average fields are then used in the quark propagators of Fig.~2.
Finally, we calculate also the light-cone momentum distributions
of the nucleons \cite{Cloet:2006bq}, and obtain the quark distributions 
in the nucleus by using the convolution formalism.

The parameters of the model are determined as usual from the 
properties of the pion and the free nucleon. In particular, the
4-Fermi coupling constants in the scalar and axial-vector diquark
channels are fitted to the mass and the axial-vector
coupling constant of a free nucleon \cite{Cloet:2005pp}. 
Then there is only one free parameter 
in the calculation of the quark distribution functions, which is the
model scale ($Q_0$) needed to perform the $Q^2$ evolution
\footnote{We use the computer code of Ref.~\cite{ev} to perform the
$Q^2$ evolution in NLO.}. By fixing
this scale to the same value as our constituent quark mass ($400$ MeV), 
we obtain a very good description of the empirical
quark distributions in the free nucleon. This is shown in Figs.~3 and 4,
where the results after the
$Q^2$ evolution (solid lines) are compared to the parametrizations 
of Refs.~\cite{Martin:2002dr,Hirai:2003pm} (dashed lines). Recently obtained results for 
the transversity distributions are shown in Ref.~\cite{Cloet:2007em}. 
The results are similar to the helicity distributions, which is 
contrary to the analysis of Ref.~\cite{Anselmino:2007fs}. Further investigations
on this point are necessary. 

\begin{figure}[tb]
\begin{center}
\begin{minipage}[t]{8 cm}
\epsfig{file=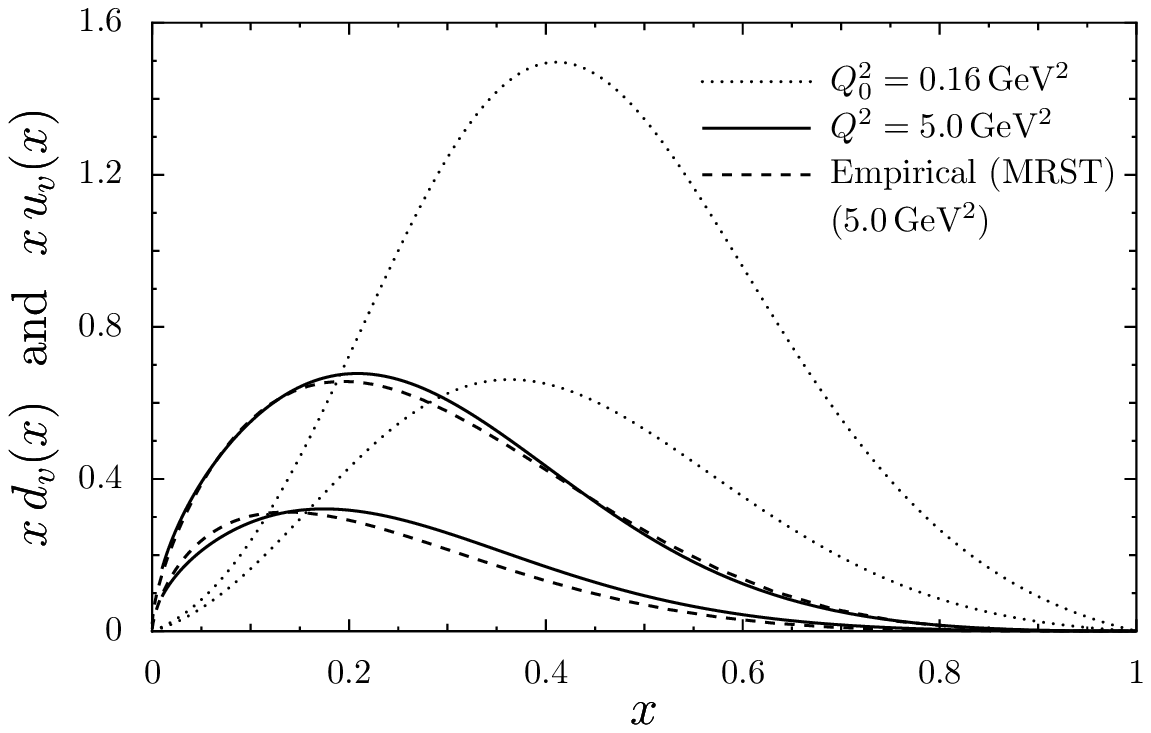,scale=0.8}
\end{minipage}
\begin{minipage}[t]{16.5 cm}
\caption{Unpolarized valence quark distributions in the free
proton. \label{fig3}}
\end{minipage}
\end{center}
\end{figure}

\begin{figure}[tb]
\begin{center}
\begin{minipage}[t]{8 cm}
\epsfig{file=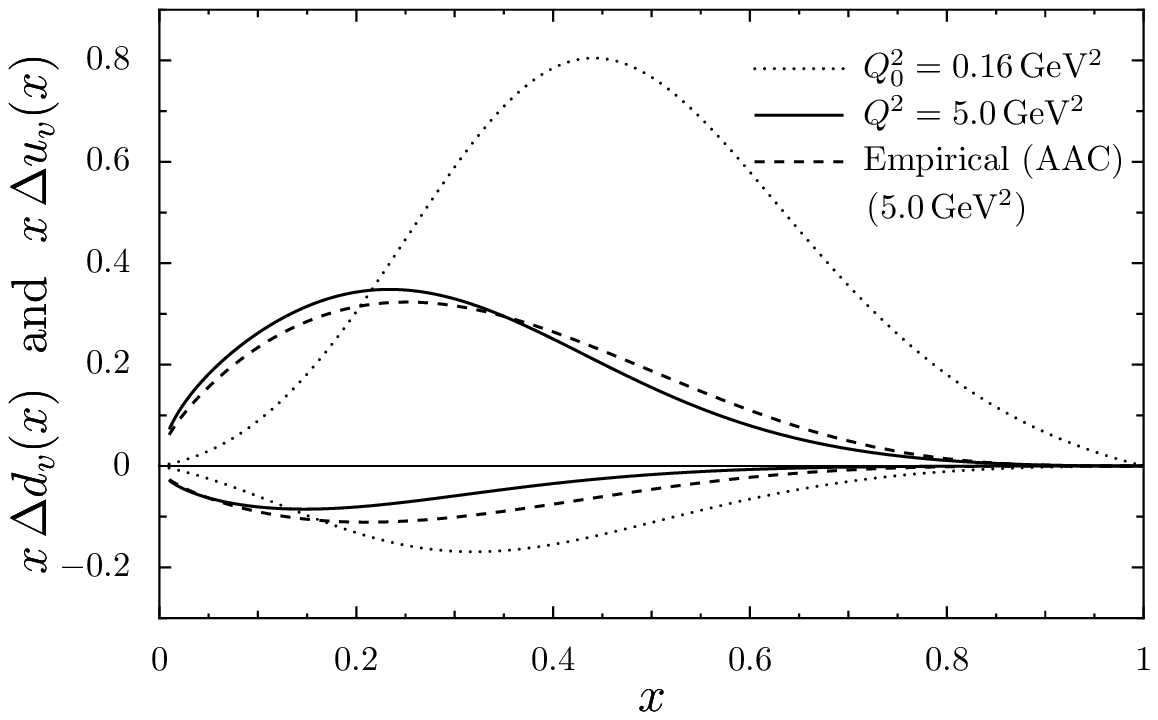,scale=0.8}
\end{minipage}
\begin{minipage}[t]{16.5 cm}
\caption{Polarized valence $u$-distributions (upper part) and 
$d$-distributions (lower part) in the free proton. \label{fig4}}
\end{minipage}
\end{center}
\end{figure}

\section{Results for nuclear quark distributions and structure functions}

In this section we will show our results for
the medium modifications of unpolarized and polarized quark
distributions and structure functions.

\begin{figure}[tb]
\begin{center}
\begin{minipage}[t]{8 cm}
\epsfig{file=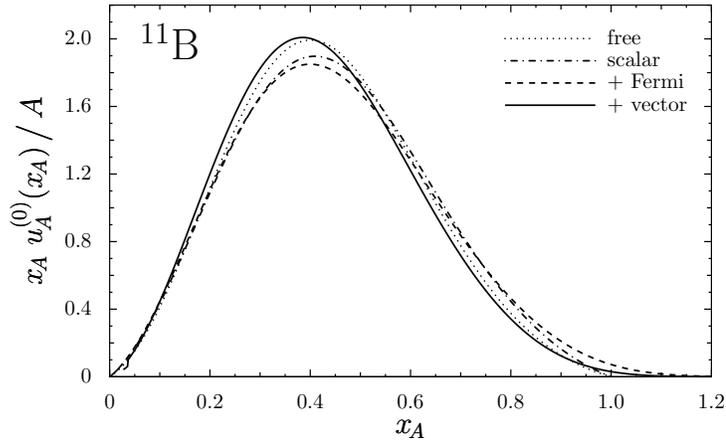,scale=0.8}
\end{minipage}
\begin{minipage}[t]{16.5 cm}
\caption{Unpolarized valence $u$-distribution in $^{11}$B. \label{fig5}}
\end{minipage}
\end{center}
\end{figure}

\begin{figure}[tb]
\begin{center}
\begin{minipage}[t]{8 cm}
\epsfig{file=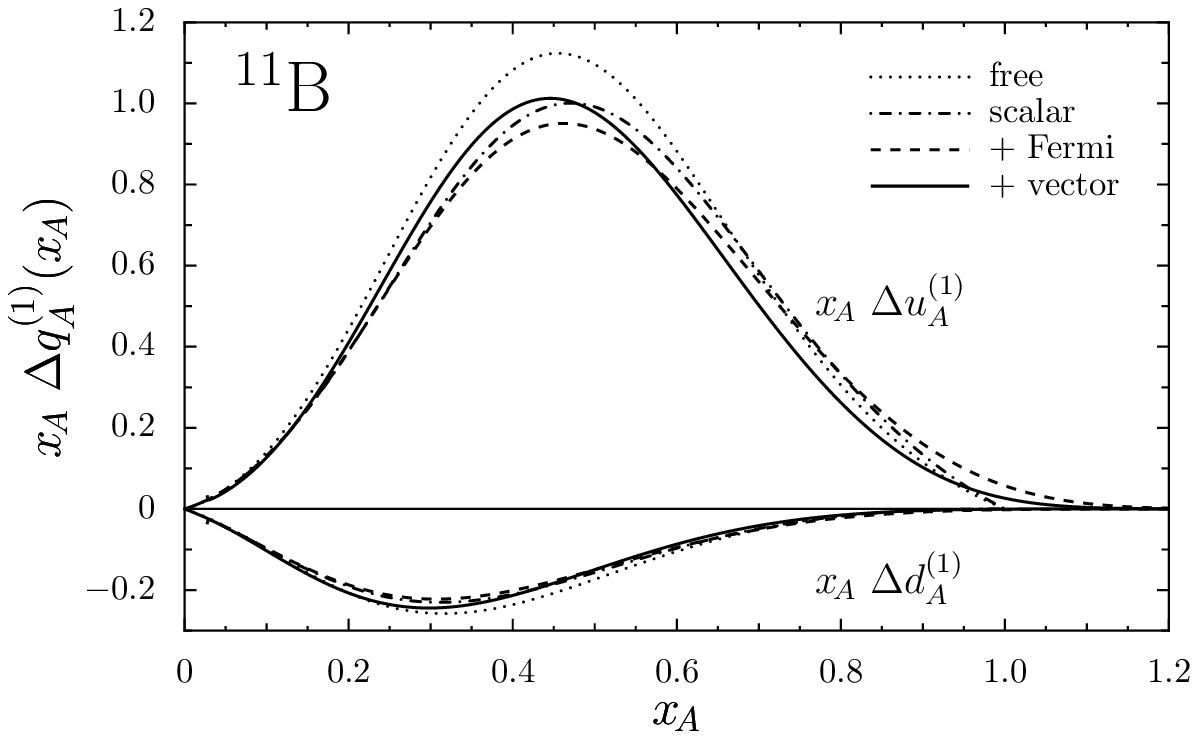,scale=0.8}
\end{minipage}
\begin{minipage}[t]{16.5 cm}
\caption{Polarized valence $u$-distribution (upper part) and $d$-distribution
(lower part) in $^{11}$B.  \label{fig6}}
\end{minipage}
\end{center}
\end{figure}

Fig.~5 shows the unpolarized valence up quark distribution in the
nucleus $^{11}$B, and Fig.~6 shows the polarized up and down quark
distributions for the same nucleus.\footnote{The distributions and structure functions shown
in Figs.~5-9 refer to the leading multipoles ($K=0$ for the
unpolarized, $K=1$ for the polarized case), which are linear
combinations of the corresponding quantities in the helicity ($H$)
basis. For details, see Ref.~\cite{Jaffe:1988up}.} The dotted lines are the free
distributions, i.e., the results obtained by neglecting Fermi motion
and medium effects. The dash-dotted lines include the effect of the
scalar mean field, the
dashed lines include further the Fermi motion, and finally the solid
lines incorporate also the effect of the vector mean field.

Comparing the dotted (free result) and the solid (full result) lines
in Fig.~5, we see that the unpolarized distribution becomes
softer in the nucleus, and that the vector potential plays a very
important role to describe this shift to smaller $x$ \cite{Mineo:2003vc,Detmold:2005cb}.  
The main features shown in Fig.~5, namely a quenching
of the distribution function at large $x$ and a small enhancement
at smaller $x$, are consistent with the EMC effect. 
On the other
hand, Fig.~6 shows that the polarized quark distributions are 
quenched in the nucleus for all values of $x$, which implies a
reduction of the quark spin sum for a bound nucleon compared to
the free nucleon case. In other words, in the medium a part of the 
quark spin is converted into orbital angular momentum. 

\begin{figure}[tb]
\begin{center}
\begin{minipage}[t]{8 cm}
\epsfig{file=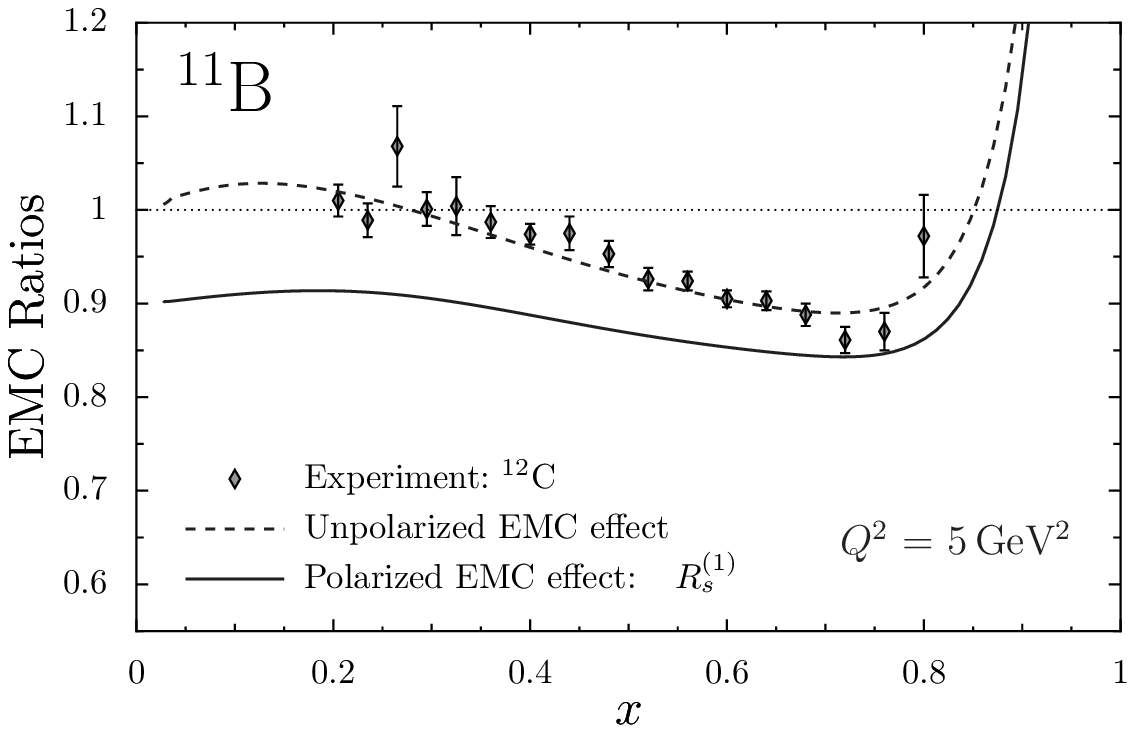,scale=0.8}
\end{minipage}
\begin{minipage}[t]{16.5 cm}
\caption{EMC ratios for $^{11}$B. The experimental data refer to
$^{12}$C. \label{fig7}}
\end{minipage}
\end{center}
\end{figure}

The resulting EMC ratios of Eq.~\eqref{ratios} for $^{11}$B are shown
in Fig.~7. It is seen that the polarized EMC effect is predicted
to be larger than the unpolarized one. As further possible 
candidates to measure the polarized EMC effect, we show the 
results for $^{7}$Li and $^{27}$Al in Figs.~8 and 9.
We see that the difference between the two EMC ratios becomes more 
pronounced as the mass number increases. 

The spin sums for the free and the bound nucleon are
listed in Table 1. Note that the quantities in the second to fifth
columns are defined by dividing out the nuclear polarization
factors from the nuclear spin sums (see Eq.~\eqref{sum}), and 
therefore directly reflect the medium modifications. The
last row shows the limit of infinite nuclear matter. We see that
the quark spin sums are appreciably
quenched in the medium. The last two columns of Table 1 show the
tensor charges for the free nucleon \cite{Cloet:2007em} and a nucleon bound in 
infinite nuclear matter.

\begin{figure}[tb]
\begin{center}
\begin{minipage}[t]{8 cm}
\epsfig{file=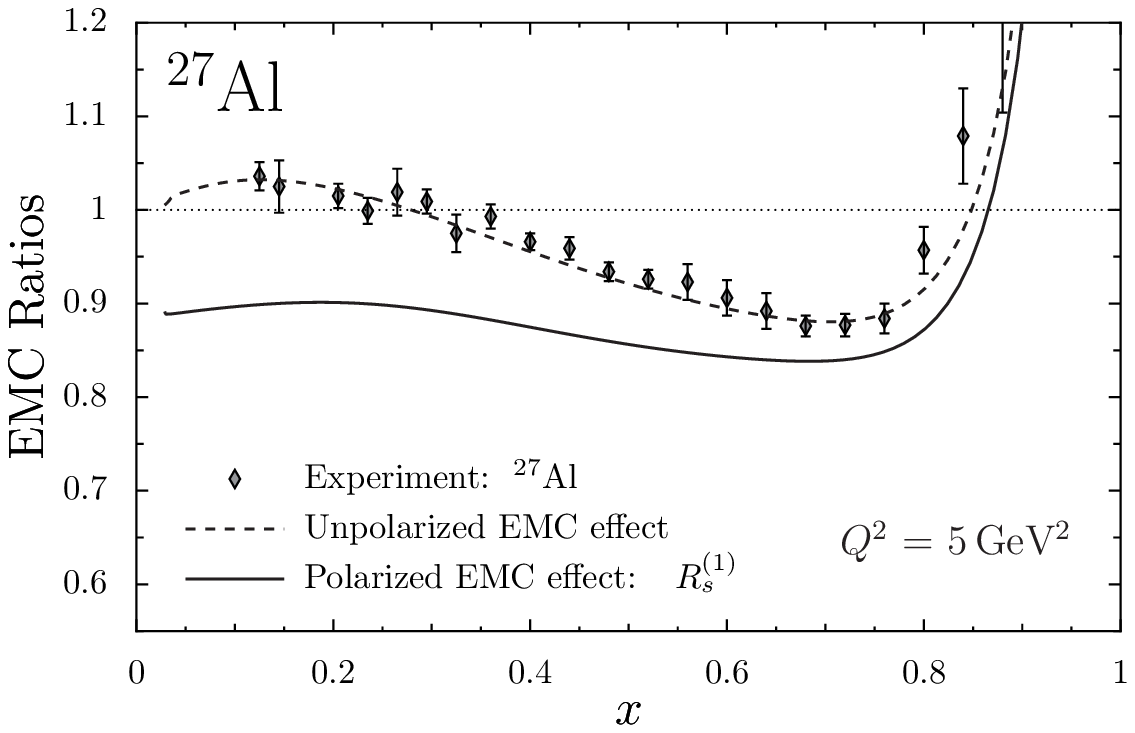,scale=0.8}
\end{minipage}
\begin{minipage}[t]{16.5 cm}
\caption{EMC ratios for $^{27}$Al. \label{fig8}}
\end{minipage}
\end{center}
\end{figure}

\begin{figure}[tb]
\begin{center}
\begin{minipage}[t]{8 cm}
\epsfig{file=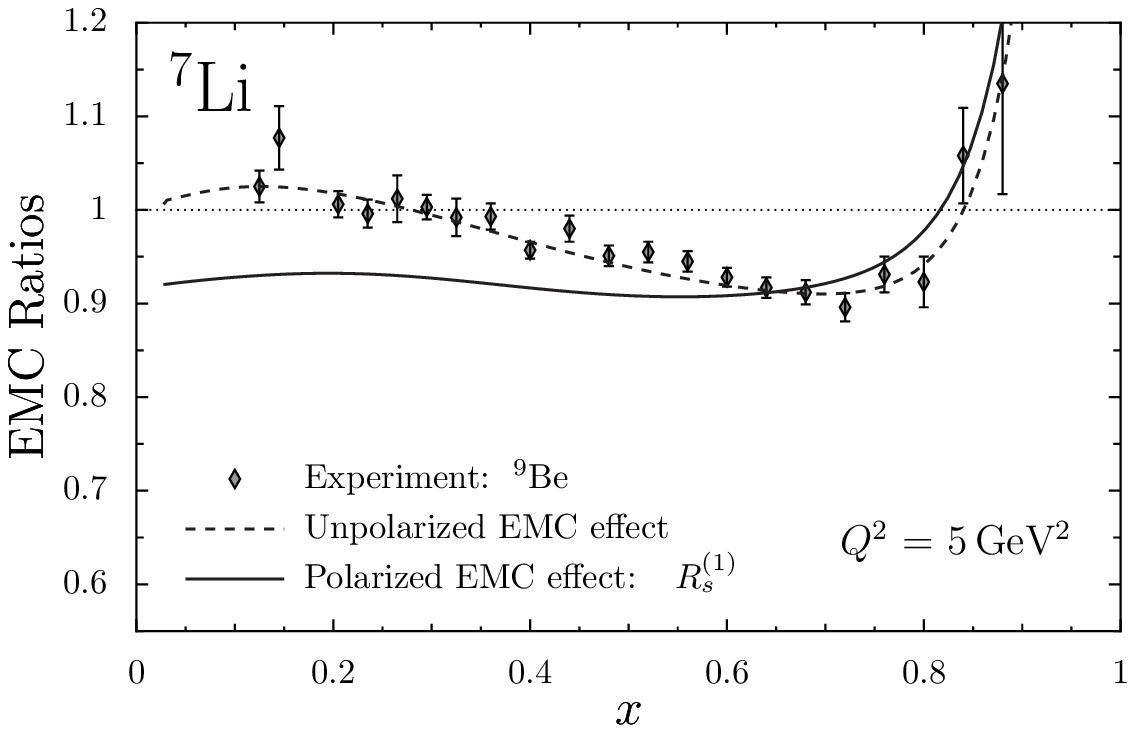,scale=0.8}
\end{minipage}
\begin{minipage}[t]{16.5 cm}
\caption{EMC ratios for $^{7}$Li. The experimental data refer to
$^{9}$Be. \label{fig9}}
\end{minipage}
\end{center}
\end{figure}

\begin{table}[tbp]
\begin{center}
\begin{minipage}[t]{16.5 cm}
\caption{Quark spin sums for a free proton and a proton bound in the
nuclear medium.}
\label{tab:obe}
\end{minipage}

\vspace{0.4 cm}

\begin{tabular}{|l|cccc|cc|}
\hline 
   & $\Delta u$ & $\Delta d$ & $\Sigma$ & $g_{A}$ & 
$\Delta_T u$ & $\Delta_T d$   \\
\hline
$p$           & 0.97  & -0.30   &  0.67    & 1.27 &  1.04 & -0.24 \\  
$^7$Li         & 0.91  & -0.29   &  0.62    & 1.19   & & \\
$^{11}$B       & 0.88  & -0.28   &  0.60    & 1.16  & & \\
$^{15}$N       & 0.87  & -0.28   &  0.59    & 1.15  & & \\
$^{27}$Al      & 0.87  & -0.28   &  0.59    & 1.15  & & \\
nucl. matt.   & 0.74  & -0.25   &  0.49    & 0.99  & 0.93  & -0.23  \\    
\hline
\end{tabular}
\end{center}
\end{table} 

\section{Extension to fragmentation functions}

Here we wish to discuss a framework to extend the
model to fragmentation functions. Numerical results for fragmentation
functions will be presented in a future publication \cite{progress}.
 
There exists a relation between the quark distribution inside a hadron
($f_q^h(x)$) and the quark fragmentation function into a hadron
($D_q^h(z)$), which is known as the Drell-Levy-Yan relation \cite{Drell:1969jm}.
Let us discuss this relation, starting directly from the operator
definitions
\begin{align}
f_q^h(x) &= \frac{1}{2}\, \hat{\sum}_n\, \delta\left(p_- x - p_- + p_{n-}\right)
\left\langle p|\overline{\psi}|p_n \right\rangle \gamma^+ \left\langle p_n|\psi|p \right\rangle 
\,, \label{fq}  \\
D_q^h(z) &= \frac{z}{6} \, \frac{1}{2}\, \hat{\sum}_n\,
\delta\left(\frac{p_-}{z}- p_- - p_{n-}\right)  \left\langle p,\overline{p_n}\left|\overline{\psi}
\right|0 \right\rangle \gamma^+  \left\langle 0|\psi|p,\overline{p_n} \right\rangle \,.   
\label{opd}
\end{align}
Here $|p \rangle$ denotes the hadron state (we assume a nucleon 
for definiteness), 
and $\hat{\sum}_n$ is the sum over
the intermediate states $|p_n \rangle $, including an integral over the
momentum and sums over spin and isospin projections. For later
convenience, the sum in Eq.~\eqref{opd} is taken over the antiparticle
states ($\overline{p_n}$). 

We can express the matrix element in the distribution function (\ref{fq})
as
\begin{equation}
\langle p_n|\psi |p \rangle = \overline{\Gamma}(p,p_n)\,\sqrt{N_p}\,
u_N({\bold p}s) \,, \label{m1p}
\end{equation}
where the Dirac matrix $\overline{\Gamma}(p,p_n)$ is the Fourier transform of
the Green function 
$ \left\langle p_n \left|\left( \psi(0) \overline{\Phi}(x)\right)\right|0 \right\rangle$ with 
$\Phi$ the 
nucleon field, and $\sqrt{N_p}$ is a normalization factor
for the nucleon spinor $u_N$. Using crossing and charge 
conjugation symmetries, the matrix element in the fragmentation function 
(\ref{opd}) can then be expressed as
\begin{eqnarray}
\langle 0|\psi|p,\overline{p_n} \rangle = - \sqrt{N_p} 
\overline{v}_N({\bold p}s) 
\Gamma(-p,p_n)\, C   \label{me}
\end{eqnarray}
where $C = i \gamma^2 \gamma_0$. Inserting these matrix elements
and their conjugates into (\ref{fq}) and (\ref{opd}), it is easy to
verify that
\begin{equation}
D_q^h(z) = \frac{-z}{6} \, f_q^h\left(x=\frac{1}{z}\right)\biggl|_{p\rightarrow -p} \,\,, 
\label{fdd}
\end{equation}
where $p\rightarrow -p$ means to reverse all 4 components of $p^{\mu}$,
and after this replacement $p^0=E_N({\bold p})>0$.

The effect of the replacement $p\rightarrow -p$ on the distribution
function Eq.~\eqref{fq} is seen most easily by expressing it in 
terms of $\Gamma$ and the quark momentum $k=p-p_n$ as follows:
\begin{align}
f_q^h(x) &= \frac{N_p}{8 M_N} \sum_n \int \frac{{\rm d}^4 k}{(2 \pi)^3}\,
\frac{\Theta(p_-(1-x))}{2 p_-(1-x)}\, \delta\left(k_+ - e_N({\bold p})+ e_n
({\bold p}- {\bold k})\right) \nonumber \\
&\hspace{30mm} \times \delta(k_- - p_- x)\ 
{\rm Tr}\left((\fslash{p}+M_N) \Gamma(p,p-k) \gamma^+ 
\overline{\Gamma}(p,p-k) \right), \nonumber \\
& \equiv  \Theta(1-x) F(x)\, . 
\label{cc}
\end{align}
Here we expressed the integrand using light-cone variables, and the
``energies'' of the intermediate state (invariant mass $M_n$) and
the nucleon are defined by 
$e_n({\bold p}_n) = ({\bold p}_{n \perp}^2 + M_n^2)/(2 p_{n-})$
and $e_N({\bold p}) = ({\bold p}_{\perp}^2 + M_N^2)/(2 p_{-})$.
From (\ref{fdd}) and (\ref{cc}) we obtain finally
\begin{equation}
D_q^h(z) = - \Theta(1-z)\, \frac{z}{6} \, F\left(x=\frac{1}{z}\right)\,.  \label{rr2} 
\end{equation}
(For spin zero bosons, there is no minus sign in Eq.~\eqref{rr2}.)
This shows that $f_q^h$ and $D_q^h$ are essentially one and the
same function, defined in different regions of the variable.
This important result allows us to extend our investigations on
the distribution functions presented in this paper to the
fragmentation functions. The numerical results and detailed
discussions will be presented in a future publication \cite{progress}.

\section*{Acknowledgements}

This work was supported by: Department of Energy, Office of Nuclear Physics,
contract no. DE-AC02-06CH11357, under which UChicago Argonne, LLC, operates
Argonne National Laboratory; contract no. DE-AC05-84ER40150, under which
JSA operates Jefferson Lab, and by the Grant in Aid for Scientific Research
of the Japanese Ministry of Education, Culture, Sports, Science
and Technology, project no. C-19540306.

\end{document}